\documentclass[12pt,preprint]{aastex}
\usepackage{graphicx}

\shorttitle{MOA-2011-BLG-293Lb: First Microlensing Planet possibly in the Habitable Zone}
\shortauthors{Marx et al.}

\begin{document}


\title{MOA-2011-BLG-293Lb: First Microlensing Planet possibly in the Habitable Zone}


\author{V. Batista\altaffilmark{1,7,8}, J.-P. Beaulieu\altaffilmark{2,8}, A. Gould\altaffilmark{1,7}, D.P. Bennett\altaffilmark{3,8,9}, J.C Yee\altaffilmark{1,7}, A. Fukui\altaffilmark{4,9}, B.S.~Gaudi \altaffilmark{1,7}, T. Sumi\altaffilmark{5,9}, A. Udalski\altaffilmark{6,10}}





\altaffiltext{1}{Department of Astronomy, Ohio State University, 140 West 18th Avenue, Columbus, OH 43210, USA; virginie@astronomy.ohio-state.edu,
gould@astronomy.ohio-state.edu, jyee@astronomy.ohio-state.edu, gaudi@astronomy.ohio-state.edu}
\altaffiltext{2}{Institut d'Astrophysique de Paris, 98Bis Boulevard Arago, 75014 Paris, France; beaulieu@iap.fr}
\altaffiltext{3}{University of Notre Dame, Department of Physics, 225 Nieuwland Science Hall, Notre Dame, IN 46556-5670, USA; bennett@nd.edu}
\altaffiltext{4}{Okayama Astrophysical Observatory, National Astronomical Observatory of Japan, Asakuchi, Okayama 719-0232, Japan;\\ afukui@oao.nao.ac.jp}
\altaffiltext{5}{Department of Earth and Space Science, Graduate School of Science, Osaka University, Toyonaka, Osaka 560-0043, Japan;\\ sumi@ess.sci.osaka-u.ac.jp}
\altaffiltext{6}{Warsaw University Observatory, Al. Ujazdowskie 4, 00-478 Warszawa, Poland; udalski@astrouw.edu.pl}
\altaffiltext{7}{Microlensing Follow Up Network ($\mu$FUN),
http://www.astronomy.ohio-state.edu/∼microfun}
\altaffiltext{8}{Probing Lensing Anomalies Network (PLANET), http://planet.iap.fr}
\altaffiltext{9}{Microlensing Observations in Astrophysics (MOA) Collaboration,\\
http://www.phys.canterbury.ac.nz/moa}
\altaffiltext{10}{Optical Gravitational Lens Experiment (OGLE) Collaboration, http://ogle.astrouw.edu.pl/}

\begin{abstract}
We used Keck adaptive optics observations to identify the first planet discovered by microlensing to lie in or near the
habitable zone, i.e., at projected separation $r_\perp=1.1\pm 0.1\,$AU from
its $M_{L}=0.86\pm 0.06\,M_\odot$ host, being the highest microlensing mass definitely identified.  The planet has a mass $m_p = 4.8\pm 0.3\,M_{\rm Jup}$, and could in
principle
have habitable moons.  This is also the first planet to be identified
as being in the Galactic bulge with good confidence: $D_L=7.72\pm 0.44$ kpc.  The planet/host
masses and distance were previously not known, but only estimated
using Bayesian priors based on a Galactic model (Yee et al. 2012).  These estimates
had suggested that the planet might be a super-Jupiter orbiting an
M dwarf, a very rare class of planets.
We obtained high-resolution $JHK$ images using Keck adaptive optics
to detect the lens and so test this hypothesis.  We clearly detect
light from a G dwarf at the position of the event, and exclude all
interpretations other than that this is the lens with high confidence ($95\%$),
using a new astrometric technique. The calibrated magnitude of the planet host star is $H_{L}=19.16\pm 0.13$. We infer the following probabilities for the three possible orbital configurations of the gas giant planet: $53\%$ to be in the habitable zone, $35\%$ to be near the habitable zone, and $12\%$ to be beyond the snow line, depending on the atmospherical conditions and the uncertainties on the semimajor axis.
\end{abstract}


\section{Introduction}

Gravitational microlensing is unique in its sensitivity to exoplanets
beyond the “snow line,” where the core accretion theory predicts that the most massive
planets will form (Lissauer 1993; Ida \& Lin 2004; Kennedy et al. 2006). It allows the exploration of host-star 
and planet populations whose mass and separation are not probed
by any other method. Statistical microlensing analyses indicate that planets beyond the snow line are more common than planets of a similar mass ratio ($q  \sim 4\times 10^{-4}$) found by
radial velocities at smaller separations (Gould et al. 2010, Cumming et al. 2008). This comparison provides strong evidence
that most giant planets do not migrate inward very far.
Since the efficiency of the microlensing
method does not depend on detecting light from the host star, it
also allows one to probe essentially all stellar types over distant 
regions of our Galaxy. In particular, microlensing is an excellent
method to explore planets around M dwarfs, which are the most common stars
in our Galaxy, but often a challenge for other techniques. Roughly half of all microlensing events toward the Galactic bulge stem from stars with mass $\leq 0.5~M_\odot$ (Gould 2000). 
The correlation between gas giant planet frequency and mass of the host star has been investigated for short orbits probed by the Doppler technique. A-type stars appear to have a significantly higher frequency of gas giant planets than solar-type stars (Bowler et al. 2010), and this occurence decreases dramatically around M dwarfs (Johnson et al. 2010, $3\%$ compared to $14\%$ for A-stars). Thus, at least for short periods, there is strong evidence that the frequency of gas giant planets increases with star mass, favoring the accretion model of planet formation (Ida \& Lin 2005, Laughlin et al. 2004). This theory has been challenged by several discoveries of gas giant planets around small stars by microlensing (e.g. Dong et al. 2009a, Udalski et al. 2005, Batista et al. 2011), as well as the large overall frequency of somewhat less massive giant planets around microlensing host stars found by Gould et al. (2010). As an alternative explanation, the
gravitational instability model predicts that gas giant planets can form
around M dwarfs with sufficiently massive protoplanetary disks (Boss et al. 2006, 2011).
\\

Microlensing thus probes a region of parameter space that provides important tests of planet formation theories. However, the robustness of the inferences about exoplanet demographics made by microlensing and thus the ability of this method to fully achieve this potential is sometimes limited by difficulties in determining the mass of the host star. The microlensing
light-curve modeling gives the planet-star
mass ratio, $q$, and the projected separation, $d$, in
units of the Einstein radius, $\theta_{\rm E}$. Most planetary
microlensing events have sharp light curve features
that reveal effects due to the finite angular size of
the source star, yielding to the measurement of $\theta_{\rm E}$.
This measurement and the so-called microlens
parallax $\pi_{\rm E}$ enable to determine directly the masses
and distance of the lens and its companion(s). Indeed
parallax effects can be detected as distortions
in the microlensing light curve due to the Earth's orbital
motion. However, about half of all planetary events
do not show detectable parallax, or in some cases,
this effect can be degenerate with the orbital motion
of the planet itself (Batista et al. 2011).
In the absence of such second order effects in the light curve of the event, the recourse of a Bayesian analysis using Galactic models is legitimately required but it tends to standardize the microlensing spectrum of detections, encouraging the M dwarf scenario and postulating strong priors on the system location in the Galaxy.
When microlensing parallax is not detectable from the lightcurve of the event itself, another alternative is to measure directly the flux from the lens using high resolution images after the event has basically ended.

The present work is based on adaptive optics (AO) observations of MOA-2011-BLG-293Lb, a planetary system whose physical properties were unclear. Our first motivation was its potential to be a new member to the population of supermassive planets orbiting M dwarfs.
We successfully measured the lens flux and this analysis rules out the M dwarf hypothesis. However, the investigation arrives at some unexpected conclusions.

It is the first microlensing planet at the edge of the habitable zone and it orbits a star similar to the Sun. We also show that this system is certainly situated in the bulge, being the furthest planet discovered to date with a very high probability of being located in this region. A gas giant planet in the bulge, such as MOA-2011-BLG-293Lb, may represent a challenge for some formation theories, as we discuss in section 6.

These surprising results highlight the importance of a systematic search for physical measurements, from the event light curve and additional high resolution observations if needed. They also suggest that we may have missed serendipitous detections of unusual systems in the past when Bayesian analyses are applied, which naturally are preconditioned to certain host stars and thus planet characteristics. Indeed, here we confirm that microlensing is sensitive to a broad variety of planetary systems, including those possibly situated in the habitable zone. It is also commonly perceived that microlensing hosts are M dwarfs, whereas here we demonstrate that some are G dwarfs, as expected.
\\

Another aspect presented in this paper is a technique to obtain accurate astrometry with adaptive optics, thanks to the unambiguous detection of excess light in addition to the source flux. Using the brightness variations of the source at two different epochs, we derive an upper limit on the separation between the position of the source and the measured excess of flux. Similar techniques have been used in the past, to distinguish the source from the blended light in microlensing events (Golberg \& Wozniak 1998), and to identify binary stars for transit events (Jenkins et al. 2010). We show that this technique can lead to substantial improvement in lens identification from high resolution (AO or space-based) imaging.
\\

\section{MOA-2011-BLG-293: a third microlensing candidate for supermassive planets around M dwarfs}

MOA-2011-BLG-293 is a microlensing planetary event observed in July 2011 and presented in Yee et al. (2012). This event was observed by all three current microlensing survey telescopes: OGLE, MOA and Wise Observatory. The planet is robustly detected in the survey+follow-up data ($\Delta \chi^2 \sim 4500$) with a mass ratio $q=5.3\pm 0.2\times 10^{-3}$ and a separation to its host star of $s=0.548\pm 0.005$ Einstein radii (Yee et al. 2012). The event was alerted as a potential high-magnification event by $\mu$FUN, and thus benefitted from an intensive follow-up (cf. Figure 1). The majority of optical observations were taken in $I$-band, but CTIO took seven images in $V$-band to measure the source color and took also $H$-band images simultaneously with the $V$ and $I$ images. 
The color and magnitude of the source have been determined from the color magnitude diagram (Yee et al. 2012): $((V-I),I)_{s,0}=(0.84,19.71)\pm(0.05,0.16)$. Such a faint source implies that systematics in the baseline data with potential errors can be comparable to the source flux itself. The baseline flux had been determined with an error of 15\% because it showed an apparent decrease in the light curve, right after the microlensing event. Additional OGLE data were taken along the year following the publication of this event, and these show that the baseline flux is actually stable, thus confirming the estimate of the source flux $f_s$ in Yee et al. (2012). The error bar on $f_s$ can therefore be reduced to 5\%.

A microlensing parallax signal has been searched in this event but was not detected, nor did this search lead to any interesting constraints.

Yee et al. (2012) derived a lower limit on the absolute magnitude of the lens (host star), $M_{I,L}\geq 3.25$, from an OGLE image at baseline, corresponding to an upper limit on the mass of $M_L\leq 1.2M_\odot$. Combined with their measurement of the Einstein radius, $\theta_{\rm E}=0.26\pm 0.02$ mas, they concluded that it could be an F/G dwarf or stellar remnant in the bulge, or a late-type star closer to the Sun.
The Bayesian probability distribution given in Yee et al. (2012) was incorrectly integrated in $\log(M)$ rather than in $M$. After correcting this error\footnote{This correction, that does not affect any other analysis than Yee et al. (2012), implies bigger error bars and different values for the lens and planet masses (old values: $M_L=0.43^{+0.27}_{-0.17}\,M_\odot$ and $m_p=2.4^{+1.5}_{-0.9}\,M_{\rm Jup}$).}, they find that the Bayesian estimate for the lens mass is $M_L=0.59^{+0.35}_{-0.29}\,M_\odot$, at a distance of $D_L=7.15\pm 0.94$ kpc, implying a planet mass $m_p= 3.3^{+1.9}_{-1.6}\,M_{\rm Jup}$.

This planetary system was thus a candidate to the population of supermassive planets around small stars, such as MOA-2009-BLG-387Lb (Batista et al. 2011) and OGLE-2005-BLG-071Lb (Dong et al. 2009a, Udalski et al. 2005). 

\begin{figure}[h!]
\begin{center}
\includegraphics[width=5in,angle=-90]{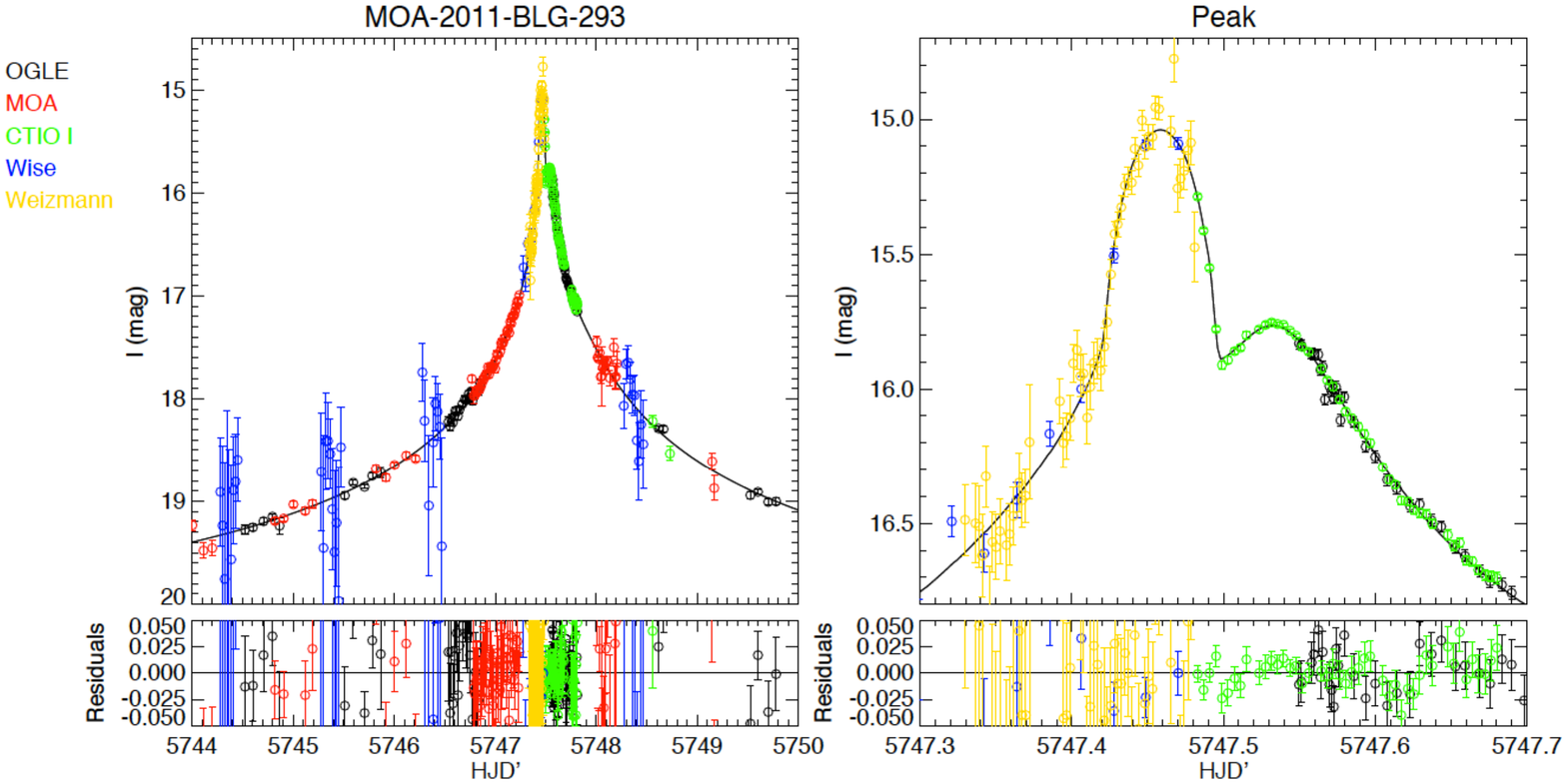}
\caption{\label{fig:f263}
Light curve of MOA-2011-BLG-293 from Yee et al. (2012). The left-hand panel shows a broad view of the light curve, while the right-hand panel highlights the peak of the event
where the planetary perturbation occurs. Data from different observatories are represented by different colors, see the legend. The black curve is the best-fit model
with a close topology ($s < 1$). The times are given in ${\rm HJD}'={\rm HJD}-2450000$.}

\end{center}
\end{figure}


\section{Adaptive optics observations}
\subsection{Going from star/planet mass ratios to physical masses}         

One of the main advantage of microlensing is that it does not depend on light from the host. But this also means that we very frequently do not measure the host light and so are unable to characterize it. The microlensing light-curve modeling 
gives the planet-star mass ratio, $q$, and the projected separation, $s$, in units of the Einstein
radius, $\theta_{\rm E}$. For most microlensing events, there is only a single measurable parameter,
the Einstein radius crossing time, $t_{\rm E}$, to constrain the lens mass, distance, and the relative
lens-source proper motion, $\mu$, where: 
\begin{equation}
\label{general}
t_{\rm E}=\frac{\theta_{\rm E}}{\mu},\qquad  \theta_{\rm E}^2=\kappa M_{L} \pi_{\rm rel},\qquad \pi_{\rm rel}={\rm AU}\left( \frac{1}{D_L}-\frac{1}{D_S}\right) \qquad {\rm and} \qquad \kappa=\frac{4G}{c^2{\rm AU}}=8.1\ {\rm mas.}M_\odot^{-1}.
\end{equation}

However, most planetary microlensing events have sharp
light curve features that reveal effects due to the finite angular size of the source star. This
allows the source radius crossing time, $t_*$, to be measured. Because the angular radius of
the source star, $\theta_*$, can be determined from its brightness and color, the measurement of $t_*$
yields $\mu = \theta_* /t_*$. Hence, for such events we are usually able to measure the product $M_{L}\pi_{\rm rel}$:
\begin{equation}
\label{pirel}
M_{L}\pi_{\rm rel}=\frac{\theta_* t_{\rm E}^2}{\kappa t_*^2},
\end{equation}

where the quantities on the right-hand-side are all observables.
It is also sometimes possible to measure
the ``microlensing parallax'', $\pi_{\rm E}$, which is an effect of Earth's orbital motion on the
microlensing light curve, and which yields a different combination of $M_{L}$ and $\pi_{\rm rel}$,
\begin{equation}
\label{pi}
\frac{\pi_{\rm rel}}{\kappa M_{L}}=\pi_{\rm E}^2.
\end{equation}

If $\theta_{\rm E}$ and $\pi_{\rm E}$ are measured, then $M_{L}$ and $\pi_{\rm rel}$ can be derived:
\begin{equation}
\label{mpirel}
M_{L}=\frac{\theta_{\rm E}}{\kappa \pi_{\rm E}} \qquad {\rm and} \qquad \pi_{\rm rel}=\theta_{\rm E}\pi_{\rm E}, 
\end{equation}
 
as was done in the case of the first triple system discovered by microlensing (Gaudi et al. 2008). However, without $\pi_{\rm E}$, we usually cannot do better than the relation in Eq. 2, which leaves one variable unconstrained. This can be resolved with high angular resolution data (to isolate the lens from its crowded field). Note that the source distance is usually quite well known, so that measuring $\pi_{\rm rel}$ is virtually equivalent to determining $D_L$.


In most cases it is possible to detect and study (or to put upper limits on) the host (lens) stars using high-resolution images, either from space or ground-based AO observations. Keck's high angular resolution
allows us to resolve the source stars from their unrelated neighbors, but the images of the
source and lens stars will generally be blended together. At the time of the microlensing
event, the lens star must be less than $1$ mas away from the source. If there are magnified $H$-band data (as in the present case), the microlensing models
determine the $H$-band brightness of the source star, so it is usually possible to determine
the $H$-band brightness of the host star (lens) by subtracting the source flux from the Keck $JHK$-band
measurement of the combined companions. Formally, this can be combined with Eq. 2 and an $JHK$-band mass-luminosity relation to yield a
unique solution for the host star mass. This would yield the planetary mass and star-planet
separation in physical units because the planet-star mass ratio, $q$, and the separation in
Einstein radius units are already known from the microlensing light curve.

\section{Adaptive optics observations of MOA-2011-BLG-293}

MOA-2011-BLG-293 was observed with the NIRC2 AO system on Keck in $H$-band with natural guide star on May 13, 2012, with the medium camera giving a plate scale of 0.02 arcsec/pixel. The exposure time was 30 seconds. We chose 5 dithering positions with a step of 2 arcseconds, and we obtained 15 good quality images. We corrected for dark current and flatfielding using standard techniques. We then performed astrometry on the 15 frames and used SWARP from the Astromatics package to stack them. Note that when using SWARP on AO images, it is important to keep in mind that the large wings of the PSF of bright star might affect
the background estimation and lead to systematic error in the photometry. During the sequence of observations, the background remained constant between the different frames, therefore we did not fit the background, and stacked the data directly. In order to measure the fluxes of the stars, we used SExtractor, in the MAG-AUTO mode with parameters that we have optimized for AO
observations. We detected 348 sources in the frame. There is no ambiguity in the identification of the source star since it is fairly isolated at the position
estimated from OGLE observations. The measured FWHM of the star is 0.12 arcsec with no evidence for nearby blend.
The Keck image is shown in Figure 2, as well as a 2MASS image of the field. The coordinates of MOA-2011-BLG-293 are (R.A., Dec)=(17:55:39.35,-28:28:36.65), $(l,b)=(1.52,-1.66)$.
\begin{figure}[h!]
\begin{center}
\includegraphics[angle=-90,width=4in]{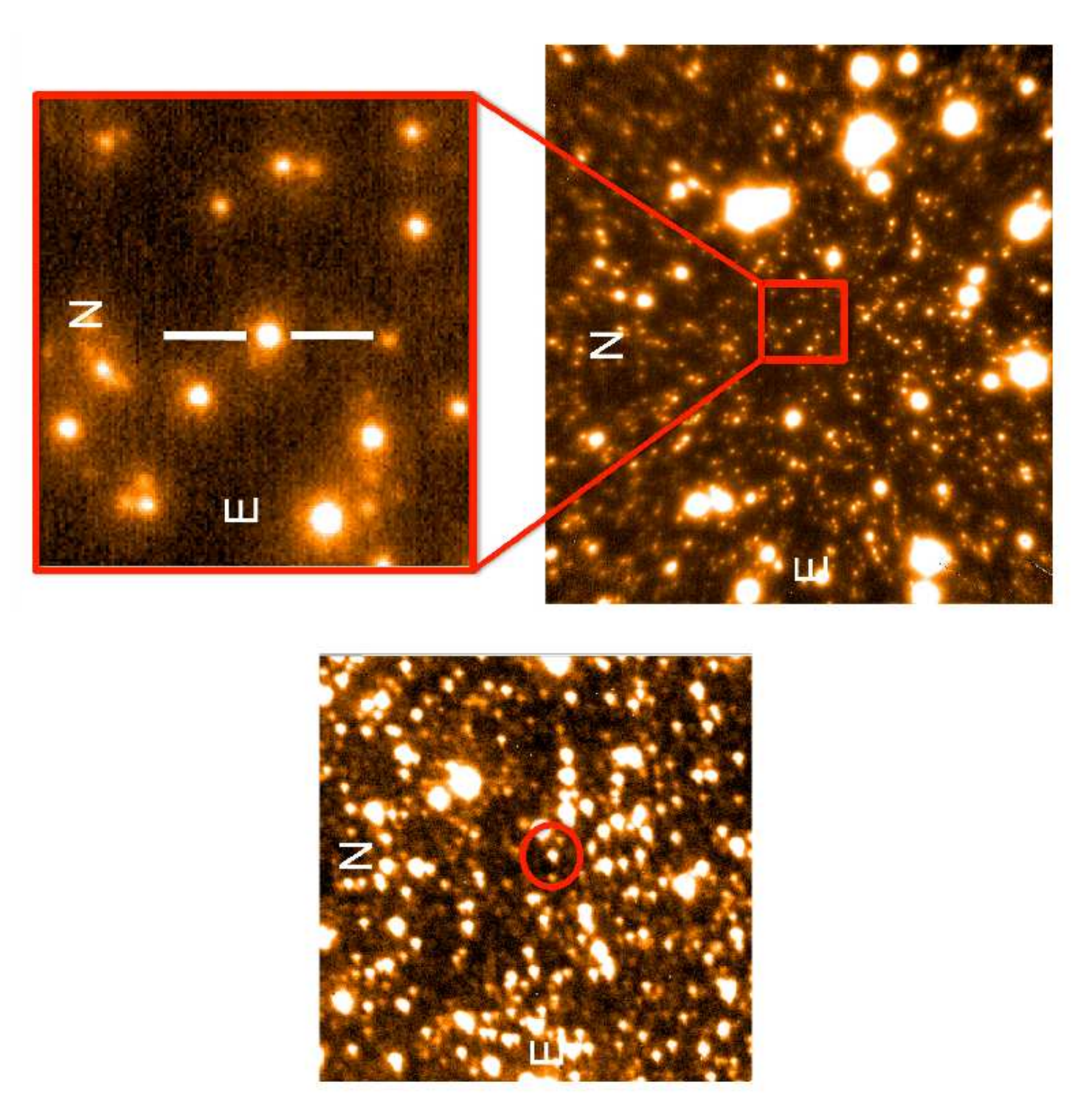}
\caption{\label{fig:fdch}
Keck image of MOA-2011-BLG-293 in $H$-band on the right ($\sim 24'' \times 21''$), zoom on the upper panel ($\sim 3.3''\times 3.3''$), and a CTIO $H$-band image of the field, when the source was magnified, on the left ($\sim 1.5'\times 1.5'$). }

\end{center}
\end{figure}

We then performed the calibration in several steps. First, we cross-identified 167 stars between CTIO $H$-band sources with the 2MASS catalog.
We excluded outliers, and calculated the calibration constant between the instrumental CTIO $H$-band and the 2MASS $H$-band to be with an uncertainty of $0.01$. Then we cross-identified 5 bright and isolated stars in both CTIO and Keck images.
We computed the calibration constant between the instrumental CTIO magnitude and the Keck instrumental magnitude with an uncertainty of $0.006$.
The resulting measured flux on the object at the position of the source is then
$H_{\rm Keck} =18.43  \pm 0.06$ in the calibrated 2MASS system.

Yee et al., (2012) have shown that $(I-H) = 0.82\pm 0.01$ (intrumental measurement), using CTIO data when the source was strongly amplified, and fitted the baseline flux of the source
to $I_{\rm s, model} = 22.27  \pm 0.05$. This corresponds in the calibrated 2MASS system to  $ H_{\rm s, model} = 19.20  \pm 0.06$. Note that the blended flux (crowded field) on CTIO images do not impact the source flux calculation that is derived from a regression over different epochs, when the source magnification varies, so that the blending flux is cancelled out.
At the target position on the Keck images, there is an excess in flux of $\Delta H = 0.77  \pm 0.09$ aligned with the source star. The Keck observations were taken almost a year after the microlensing event, so the measured flux corresponds to the baseline flux when the source is no longer magnified. According to the event timescale given by Yee et al. (2012), $t_{\rm E}=21.67$ days, the lens and the source should be separated by $t_{\rm elapsed}/t_{\rm E}\times \theta_{\rm E} \sim 14\, \theta_{\rm E}\sim 3.6$ mas.
Assuming that all this excess flux is coming from the lens star:  $ H_{L} = 19.16  \pm 0.13$.


\section{Astrometry on Keck images: how to accurately identify the lens?}
\subsection{Upper limit on the separation between the source and the excess flux}

Since the Keck images were obtained when the source was no longer magnified (almost a year after the microlensing event), and knowing that the unmagnified source and excess of flux have a comparable brightness, we can measure (or limit) the deviation in the centroid position between the Keck images and an OGLE image taken during the microlensing event, while the source was magnified and the blending negligible. This will constrain the separation between these two stars. Therefore, we align the astrometry of the Keck $H$-band image on the OGLE astrometry. Note that the target centroid on the OGLE difference images will correspond to the source position, while the centroid on Keck images is the sum of the source and the blending (excess flux). We then compute a linear transformation between the OGLE and the Keck frames considering 20 common stars in these two fields. We exclude five outliers and a visual check confirmed that these were elongated and/or far from the center of the image on AO observations. \\
Figure 3 presents the Keck and OGLE fields before and after the linear transformation. The target is represented by the crosses. The lower panel shows the (2-D) residuals in pixels for each star after the transformation.\\

We find an offset between the OGLE source and the transformed Keck "baseline object'' of $(\Delta_{\rm RA},\Delta_{\rm Dec}) = (11,14)$ mas.  There
are two distinct but related sources of error for this measurement.
First, even given perfect information about the positions of these
two objects on OGLE and Keck images, there would be an error due
to the uncertainty in the transformation.  Since the target is
approximately in the center of the reference-star system, this error
is simply given by $(\rm rms/\sqrt{\rm dof})_{\rm RA,\rm Dec} = (5,6)$ mas.  Second, the positions
of these two objects are not perfectly known.  If we assumed that
the errors in these positions were similar to those of the reference
stars, then we would have an empirical measure of this error, namely
$(\rm rms)_{\rm RA, \rm Dec} = (17,21)$ mas.  But in fact these
numbers are upper limits.  For the reference stars, the position errors
are dominated by OGLE astrometry because the stars are not fully isolated
on the OGLE image.  However, the source star is completely isolated on
the difference image, making its measurement much more precise.  We
estimate this precision to be 0.03 pixels, or 8 mas, based on the scatter
of measurements from individual difference images.  We assume that the error in the Keck position is much
smaller.  Hence, we finally adopt an error of 10 mas in each direction
for the offset.  The probability of measuring an offset as large as the
one observed, assuming that the two positions were coincident is
$\exp(-(1.1^2 + 1.4^2)/2) = 20\%$, which is not significant.  Instead,
we place a conservative upper limit on the separation, $\Delta\theta < 30$
mas. Since the source and the excess flux are very similar in $H$-band (19.20 vs 19.16), this upper limit will result in a maximum angular separation between the source and the excess flux of $d<60$ mas. This separation is small and makes the excess of light very likely due to the lens. The deviation due to the source-lens proper motion can be neglected here ($\mu=4.3 \pm 0.3 \rm{\ mas/yr}$, Yee et al. 2012). This astrometric estimate reduces the upper limit on the separation between the source and this excess flux by a factor of 2, since, without such a transformation, the only constraint for this flux would have been for it to lie within the 112 mas FWHM of the source. The identification of the excess flux nature in the next sections is subsequently facilitated by this new constraint. 

\begin{figure}[h!]
\begin{center}
\includegraphics[width=4in]{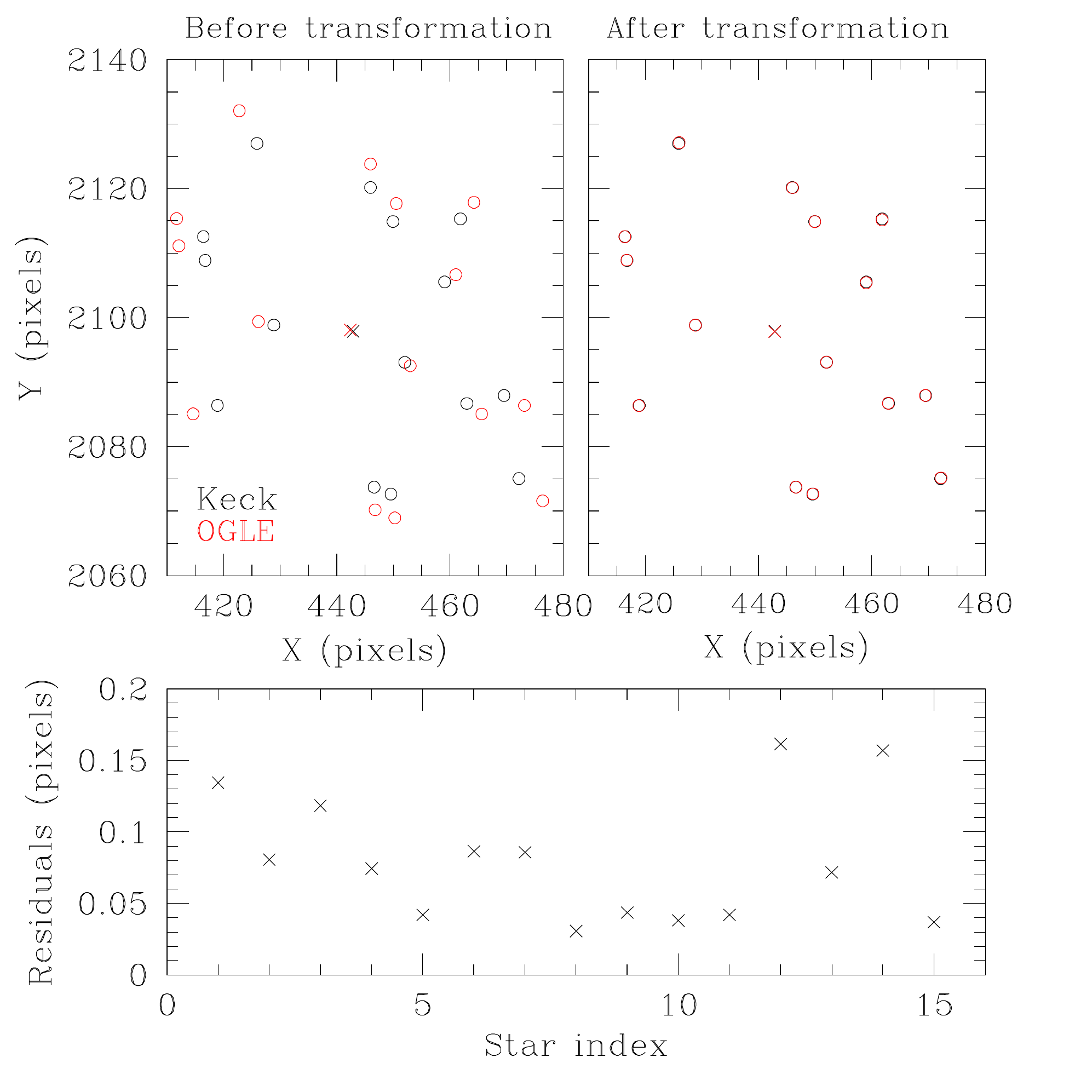}
\caption{\label{fig:transfo}
Calibration of the Keck stars coordinates with the OGLE catalog, plotted in the OGLE frame ($0.005'$/pixel). A first arbitrary transformation (scale and origin offset) has been applied to the Keck coordinates in the left panel, in order to plot the two fields in the same frame. The right panel shows the alignment of KECK and OGLE stars after the transformation. The target is represented by the crosses. The lower panel shows the (2-D) residuals after the linear transformation for each star. }

\end{center}
\end{figure}



\section{Nature of the excess flux}

Measurements of the MOA-2011-BLG-293 target flux with the AO technique clearly revealed an excess of flux in addition to the unmagnified source flux. As suggested previously, this flux is likely to be due to the lens, but to be rigorous, we must estimate the probability that it is due (or partly due) either to an ambient star on the line of sight, to a companion to the source, or to a companion to the lens. We follow a similar approach to the one presented in Janczak et al. (2010), but reaching a 4 times better precision by performing astrometric measurements.
For each ``alternative'' (ambient, source companion, lens companion) we
seek to evaluate the relative probability that the light is due to the
alternative relative to the lens, given the fact that a star of this
magnitude has been detected within 60 mas of the source.  For the
source-companion
case, both possibilities are bulge stars, so there are no additional
factors due to host population.  This is also basically true of
ambient-star
case, since most such stars are in the bulge.  However, if the excess
flux is a companion to the lens, we must consider possible lenses of
a range of masses, distributed along the line of sight.

\subsection{Ambient star}

According to our estimate of the upper separation between the two stars, the excess flux must lie within 60 mas of the source. The measured excess flux being $ H_{\rm excess} = 19.16  \pm 0.13$, we search for stars whose brightness is within $ H_{\rm excess} \pm 0.5$ mag in the field ($> 3\sigma$ to be conservative). The number of such stars is equal to 78 on the Keck image in $H$-band, so with a FOV of $27.08''\times  27.12''$, this implies a density of $0.106$  $\rm {  arcsec}^{-2}$. The probability for a star of this magnitude to lie within the 60 mas astrometric limit from the source is then equal to $0.12\%$.

\subsection{Companion to the source}

The source has a magnitude $H=19.20$ and is almost certainly a G dwarf. If the excess flux is a companion to the source, of a similar magnitude (19.16 in $H$-band), we must calculate the probability of having a binary system composed by G dwarfs for a given range of separations. The upper limit on the mass ratio between the potential companion and the source is $q\leq 1$ since the source is already at the turn-off limit, so a $q=1.1$ mass ratio would render the companion a giant. For the lower limit on $q$, we consider a lower limit on the excess flux, 0.5 mag fainter than the measured one ($> 3\sigma$ to be conservative), $H_{\rm max}=19.7$. At 8 kpc, the corresponding absolute magnitude of such an object would be about 4.6. From the isochrones given by An et al. (2007) for main sequence stars, this magnitude is associated to an approximate mass of $0.75 M_\odot$, giving a mass ratio around 0.85, if we assume that the source has an absolute magnitude of $\sim4.1$ and then a mass of $0.88M_\odot$. We will consider a slightly larger range: $0.8\leq q \leq 1.0$. According to the Table 7 in Duquennoy \& Mayor (1991) (DM91 hereafter), only $9$\% of their sample of 164 stars have companions within this range. If we restrict the range of angular separations, $s$, to our observational constraints, a maximum separation will be given by the previous assessment of 60 mas, and a minimum separation will be given by a limit below which the companion of the source would have created a bump on the light curve, i.e. $s\theta_{\rm E} \geq 1/4 \theta_{\rm E} \sim 0.065$ mas as a conservative threshold. Assuming these limits are those of semi-major axes, i.e. 0.52 AU to 480 AU at 8 kpc, we can convert them into orbital periods of the companion: $2 \leq \log P ({\rm days}) \leq 6.4$. This range represents $35 \%$ of the sample on Figure 7 of DM91, so the final probability for the source to have such a companion is $3.1 \%$.

\subsection{Companion to the lens}            

For the case that the excess light is due to a companion to the lens
rather than the source, there are again restrictions on the minimum
and maximum separation, but the analysis is complicated by the fact
that the system could be at a range of distances.
The upper limit on the angular separation of the lens and companion
is exactly the same as in Section 6.2: 60 mas.  To determine the lower
limit,
we note that if the lens had another companion, in addition to the detected planet in Yee et al. (2012), it would induce a shear on the lens gravitational field that would generate a caustic at the center of magnification of the lensing system (Chang \& Refsdal, 1979, 1984). This would create some spikes in the light curve, whose magnitudes are strongly limited by the residuals shown in Figure 1. The full width of the induced caustic would have the following expression:
$$w=4\frac{q}{s^2}$$
where $q$ is defined here as the mass ratio between the hypothetical companion and the lens.
We adopt a conservative upper limit for the shear: $\gamma = \frac{q}{s^2}<10^{-3}$. As a comparison, the caustic width of the discussed planet in Yee et al. (2012) is equal to $7\times 10^{-2}$. 
If the excess of flux comes from a companion to the lens, then the companion might be a G dwarf while the lens might be an M dwarf, resulting in a mass ratio $q \sim 2$. In this configuration, and considering the shear effect, the separation between them to avoid spikes in the light curve must satisfy: $s\geq 45$, i.e., knowing $\theta_{\rm E}=0.26$ mas, $\Delta \theta > 11$ mas, where $\Delta \theta$ is the angular separation between the lens and this second companion. 
Hence, 11 mas $< \Delta \theta <$ 60 mas, i.e. 0.74 decades in separation,
which corresponds to just over 1 decade in period.  Now, the exact
probability of a companion lying in this decade depends in principle
on the lens mass, and hence its distance.
However, because the DM91 companion distribution is
nearly flat in log period, it does not depend much.  We therefore
adopt the value $M=0.5\,M_\odot$ for the lens, which corresponds
to $D_L = 7\,$kpc and so $5.4 < \log (P/{\rm day}) < 6.5$.  Only a fraction
10\% of the 164 stars observed by DM91 had companions in this interval.
(Since this is near the gentle peak in the DM91 distribution, other
lens masses will generally lead to slightly lower probabilities.)

As just evaluated, for lens masses $M=0.5\,M_\odot$, the lens is still
in the bulge, albeit a region of lower density, but for lower masses,
it must reside in much lower density regions.  An integral over all
viable M dwarf possibilities is therefore well approximated to
the interval $0.4<q_{\rm lens}<0.6$, where $q_{\rm lens}= 1/q$ and $1.6<q<2.5$ (M dwarf with a G dwarf companion). In this range, according to Table 7 of DM91, $18.2\%$ of their sample (of 164 stars) have such companions. If we limit the period to the previous range, the final estimate probability is $2\%$ (cf. Figure 7 of DM91, since $10\%$ of the sample are within this period range).

\subsection{More complicated combination}
In the previous sections, we ignored the more complex possibility that the excess flux is coming from both the lens and an additional star along the line of sight. However, by having considered above that an additional star would be responsible for the $ H_{\rm excess} \pm 0.5$ mag, we overestimated the resulting probability and this overestimation (probability that the lens or ``other star'' each have comparable brightness to the excess flux) is roughly equal (and of opposite sign) to the ignored possibility that they are both comparable to half the excess flux. Since both these quantities are small and approximately cancel, we ignore them. Thus, we assume that the final probability for the flux not to be only due to the lens still does not exceed 5\%.

\subsection{The excess flux is due to the lens}            

This scenario is the most likely, with a $\sim 95\%$ probability according to the previous calculations. 
The measured magnitude in $H$-band is converted into absolute magnitude by:\\
\\
$$M_{\rm H}=H_{\rm L}-A_{\rm H}-{\rm DM}=H_{\rm L}-A_{\rm H}-5\log\frac{D_{\rm L}}{10\rm pc}$$\\
where $A_H$ is the extinction along the line of sight and DM the distance modulus. The Galactic coordinates of MOA-2011-BLG-293 being $(l,b)=(1.52,-1.66)$, we obtain a total extinction in $H$-band: $A_{\rm H,C}=0.65 \pm 0.12$ (Cardelli et al. 1989) or $A_{\rm H,N}=0.47 \pm 0.10$ (Nishiyama et al. 2009) (see also Gonzalez et al. 2011). Assuming a linear expression for the extinction along the line of sight, we obtain the allowed range for the lens flux in $H$-band shown in Figure 4, as a function of the system distance, where the two black solid lines correspond to the two extinction laws. The dashed curves incorporate the uncertainties on all the parameters. The excess flux detected by Keck AO observations then involves a star whose absolute magnitude is within the dashed curves (only $\sim 5\%$ chance to be below the lower limit).  

We want to correlate these measurements with a calibrated population of main sequence stars. To do so, we adopt isochrones from An et al. (2007) that provide a mass-luminosity function for different ages and metallicities of main sequence stars. We take the oldest population of 4 Gyrs and the following range for metallicity: $0.0 \leq [Fe/H] \leq +0.3$. We place them on the luminosity-distance plot by using the mass-distance relation derived from the microlensing light curve (see Eq. 1, where $\theta_{\rm E}=0.26$ mas). We then estimate the lens distance as the intersection between the flux measurements and the isochrones. This determination assumes that we know the source distance (Eq. 1). For microlensing events, the source is most likely to be situated in the bulge and we usually consider the mean distance of the bulge at the Galactic coordinates of the event. However in the present case, the lens is also possibly in the bulge, and as a consequence, small variations of the source distance could imply substantial variations in the lens distance estimate. Thus, we need to apply a distance probability profile for the source, according to the distribution of stars in the Galactic bulge. At these Galactic coordinates, the distance modulus is ${\rm DM}=-5\log(D_*/10\rm(pc))=14.520$ with a dispersion of $0.187$ (Nataf et al. 2013). We take ten source distances evenly sampled on a $2\sigma$ range around the mean. To each position of the source corresponds a position of the isochrones on the luminosity-distance plot that yields to an estimate of the lens distance. We combine the likelihood of both the source and the lens to obtain the configuration with the highest probability. The figure 4 shows this final configuration only, with the corresponding set of isochrones. An additional isochrone from Girardi et al. (2002) is shown with a 10 Gyrs population and a metallicity of $[Fe/H] = 0.0$. As a reference, the results from the Yee et al. (2012) Bayesian analysis (after correction) are shown in the shaded box, centered on $D_L=7.15 \pm 0.94$ kpc. The intersection between the isochrones and the Keck detection gives an estimate of the lens absolute magnitude: a $M_H\sim 4.2$ magnitude at a distance $D_L= 7.72\pm 0.44$ kpc, shown in the dark shaded box, for a source distance $D_S=8.35\pm 0.49$ kpc. The error bar is the quadratic sum of the uncertainties from the isochrones intersection with the Keck measurements and the calculated standard deviation of the source/lens distances.
\\
The mass can be deduced from the distance by the following expression:
\\
$$M_L=\frac{\theta_{\rm E}^2}{\kappa \pi_{\rm rel}}= 0.86 \pm 0.06\,M_\odot$$
where $\theta_{\rm E}=0.26\pm 0.02$ mas (Yee et al. 2012). 
According to the Keck measurements, the lens is then more likely to be a late G dwarf than an M dwarf. With a estimated mass of $M_L= 0.86\pm 0.06\, M_\odot$, the planet will have a mass of $m_p= 4.8\pm 0.3\,M_{\rm Jup}$. This planet is also the furthest of the discovered planetary systems so far, with the highest confidence of being situated in the bulge, i.e. within a very different stellar population (Type II versus Type I), where the stars have a higher metallicity and are likely to have formed differently than in the disk. The formation of gas giant planets in the bulge has been questioned by Thompson (2013). Indeed, if the core accretion theory is preferred, the temperature in such a dense region of stars, with a high exposure to radiation, might exceed the ice line temperature ($T_{\rm ice}\sim 150-170$ K) during a time comparable to the planet formation timescale. However, detecting at least one planet in the bulge within the sample of 21 planets discovered by microlensing suggests that such gas giant planets might form in this region anyway. 
\\

Considering the separation obtained by Yee et al. (2012) from their best fit model, $s=0.458\pm 0.005$, the distance estimate provided by the Keck measurements, $D_L=7.72\pm 0.44$ kpc, results in a projected separation $r_\perp = 1.10\pm 0.1$ AU between the planet and its star, which is within the snow line ($\sim 1.7-2.2$ AU), i.e. the location at which water
sublimated in the midplane of the protoplanetary disk (T = 170 K according to e.g. Sasselov \& Lecar 2000).
At such a distance from its star there is a possibility for the planet to be situated in the habitable zone (HZ).
Guillot et al. (1996) gives an expression of equilibrium effective temperature of the planet $T_{\rm eq}$ as a function of the
star parameters ($T_{\rm eff},\, R$) and albedo $A$ assuming that the heat is uniformly distributed:

$$T_{\rm eq}= T_{\rm eff}   (R_*/(2a))^{1/2}(1-A)^{1/4}$$

The lens star having a mass $M_L= 0.86 M_\odot$, we take the hypothesis of solar abundance and derive a temperature $T_{\rm eff} = 5180$ K from the isochrones of An et al. (2007). We derive a radius of $R_*=0.82\, R_\odot$ from the Stefan-Bolztmann law. As we are more interested in knowing if an hypothetical satellite of this gaseous planet would be habitable, we assume an albedo in the range [0.2, 0.7] according to the solar system population. Indeed, although the data do not explicitely show any signature of a companion to the Jupiter planet, this possibility is not ruled out.
As a consequence, we estimate the equilibrium temperature of the planet and its hypothetical companions to be $159\,K<T_{\rm eq}<204\,K$.
The planet is apparently at the edge between the snow line and the habitable zone, but considering a potential greenhouse warming effect, the surface temperature of a possible companion can be suitable for habitability. Indeed, if we apply the expression (2) and (3) of Selsis et al. (2008), for three different atmospheric compositions, we obtain the following ranges for the habitable zone (assuming $L/L_\odot = 0.434$ from the isochrones): $0.56-1.14$ AU ($0\%$ cloud), $0.45-1.32$ AU ($50\%$ clouds) and $0.30-1.61$ AU ($100\%$ clouds). In $\sim 60\%$ of the atmospheric configurations (interpolation of the clouds density vs radius by a second order polynomial), if we consider a random orbit $a\sim 1.15\times r_\perp \sim 1.27$ AU
\footnote{The 1.15 factor comes from the assumption that, for circular orbits, the distribution of
$\cos(\beta)$ is uniform, where $r_\perp=a*\sin(\beta)$, and $a$ is the semimajor axis.
Thus, the median of $\cos(\beta)$ is 0.5. Therefore, $<a/r_\perp>_{\rm med} = 1/\sqrt(1-0.5^2)\sim 1.15$.}, MOA-2011-BLG-293Lb is situated in the habitable zone according to their criteria. 
More recent models for habitability 
around main-sequence stars are given by Kopparapu et al. (2013) and using their online HZ calculator confirms these conclusions on MOA-2011-BLG-293, placing its 
empirical habitable zone between 0.52 and 1.28 AU, taking Mars and Venus as references for outer and inner HZ.

Moreover, a recent definition of exomoon habitability was given by Heller \& Barnes (2013) (hereafter HB13) which takes into account the illumination from the planet and tidal heating in the satellite, in addition to the direct heating from the star (see Eq. 22 in HB13). Applied to MOA-2011-BLG-293Lb, these additional energy sources, and in particular the tidal heating, could substantially increase the amount of energy available to a hypothetical companion and thereby compensate for the weak stellar irradiation. As an illustration, the stellar flux received by MOA-2011-BLG-293Lb at 1.1 AU is around $86\, {\rm W/m}^2$, but if a hypothetical exomoon at a distance of about 6-7 planetary radii from the planet would be in a slightly eccentric orbit, then the global average flux could be as high as $300\, {\rm W/m}^2$, i.e. more than the solar flux absorbed by the Earth and even close to the moon's runaway greenhouse limit.
\\

Although the wide orbit model is disfavored in Yee et al. (2012) analysis compared to the one we adopted above ($\Delta \chi ^2=3.9$, i.e. a probability of $12\%$), we have to consider this other possible configuration, , $s=1.83\pm 0.01$, giving a projected separation of $r_\perp = 3.67$ AU, i.e. a cold gas giant planet beyond the snow line and outside of the habitable zone.

The final probability for MOA-2011-BLG-293Lb to have a close orbit ($P= 88\%$) and to be situated in the habitable zone ($P\simeq 60\%$) is then equal to $53\%$. The close orbit solution would generate a RV signal of $\sim 100\, {\rm m.s}^{-1}$ and could probably be detected with the next generation telescopes (e.g. instrument CODEX of E-ELT, Pasquini et al. 2010).

\begin{figure}[h!]
\begin{center}
\includegraphics[width=4in]{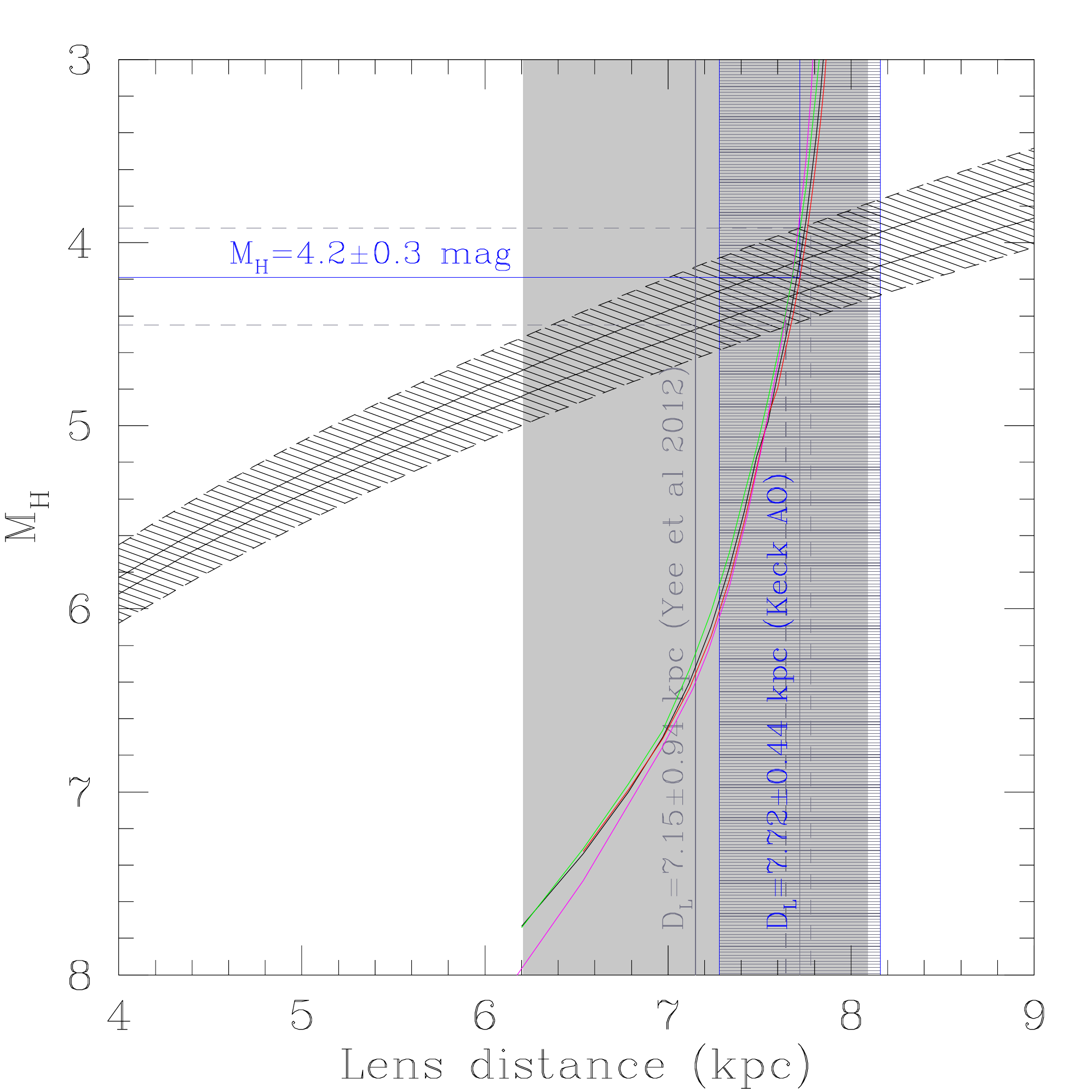}
\caption{\label{fig:mag}
Estimate of the lens absolute magnitude measured by Keck AO observations in $H$-band (black curves, dashed area including error bars). The probability that the lens absolute magnitude be below the lower limit (dashed lower curve) is less than 5\%. The colored curves are 4 Gyrs isochrones for main sequence stars from An et al (2007), with the following metallicities: $[Fe/H]=0.0$ (dark blue), $[Fe/H]=0.2$ (green), and, $[Fe/H]=0.3$ (red). An additional isochrone from Girardi et al. (2002) is shown in magenta with a 10 Gyrs population and a metallicity of $[Fe/H] = 0.0$. The previous lens distance estimate given by Yee et al. (2012) is shown in the shaded box (after correction). The intersection between the keck measurements and the isochrones define the allowed area for the lens parameters shown between the grey dashed lines. The final lens distance estimate is shown in the dark shaded box}

\end{center}
\end{figure}

\section{Conclusion}

Microlensing is complementary to the radial velocity technique in that it is
sensitive to planets with larger semimajor axes, closer to their
supposed birth sites, and the only way to find cold planets orbiting M dwarfs. Indeed, based on the analysis of 13 well-monitored
high-magnification events with 6 detected planets, Gould et al. (2010)
found that the frequency of gas giant planets at separations of $\sim
2.5~{\rm AU}$ orbiting $\sim 0.5~M_\odot$ hosts was quite high and, in
particular, consistent with the extrapolation of the frequencies of
small-separation gas giant planets orbiting solar mass hosts inferred from
radial velocity surveys out to the separations where microlensing is 
most sensitive.
This suggests that low-mass stars may form gas giant planets as efficiently as do
higher mass stars, but that these planets do not migrate as
efficiently.
Such an efficient gas giant planet formation may help the explanation of the many free-floating planet candidates found by Sumi et al. (2011).

Current and future microlensing surveys are particularly sensitive to
large $q$ planets orbiting M dwarf hosts, for several reasons. As
with other techniques, microlensing is more sensitive to planets with
higher $q$. In addition, as the mass ratio increases, a larger
fraction of systems induce an important subclass of resonant-caustic
lenses. Resonant caustics are created when the planet happens to have
a projected separation close to the Einstein radius of the
primary and increase subsequently the microlensing detection efficiency.

Here we used AO observations with the Keck instrument to obtain new constraints on the host mass of a super-Jupiter MOA-2011-BLG-293Lb. 
We have published a number of high angular resolution images of microlensing events using the HST, VLT and Keck instruments (Janczak et al. 2010, 
Batista et al. 2011, Kubas et al. 2011, Muraki et al. 2011, Sumi et al. 2010, Bennett et al. 2006, Dong et al. 2009). 
On the case we consider here, MOA-2011-BLG-293Lb, the target was easy to isolate from its neighbor stars. We detected an excess of flux in addition to the source flux and were able to perform accurate relative astrometry on the target position. This enabled us to determine the nature of the host star as being a late G dwarf rather than an M dwarf, with a mass $M_L=0.86\pm 0.06\,M_\odot$, which is still within the range given by Yee et al. (2012) from their Bayesian analysis, but very close to the upper limit. The planet is thus a massive Jupiter of $m_p=4.8\pm0.3\,M_{\rm Jup}$ separated from its host star by $r_\perp \sim 1.10$ AU. This configuration puts the planet at the edge of the habitable zone, with a probability of $53\%$ to be situated in the HZ. Plus, this is the furthest planet ever discovered, with a very high confidence of being situated in the bulge. 

These unexpected properties of the lens highlight the importance of a systematic search for physical measurements from additional high resolution observations in the absence of second order effects in the microlensing light curve, such as "microlensing parallax". In such cases, the recourse of a Bayesian analysis using Galactic models is legitimately required but tends to standardize the microlensing spectrum of detections, encouraging the M dwarf scenario and postulating strong priors on the system location in the Galaxy. Other ongoing analyses of AO observations of microlensing events (Beaulieu et al. in prep, Bennett et al. in prep, Fukui et al. in prep) will also confrontate their results to previous Bayesian analyses. More generally, the ongoing and future AO campaigns will show if such an offset from the Bayesian estimate is common.



\acknowledgments
  This work was supported by a NASA Keck PI Data Award, administered by the NASA
Exoplanet Science Institute. Data presented herein were obtained at the W. M. Keck Observatory
from telescope time allocated to the National Aeronautics and Space Administration through the
agency’s scientific partnership with the California Institute of Technology and the University of
California. The Observatory was made possible by the generous financial support of the W. M.
Keck Foundation.
AG and BSG were supported by NSF grant AST 1103471. The OGLE project has received funding from the European Research Council
under the European Community's Seventh Framework Programme
(FP7/2007-2013) / ERC grant agreement no. 246678 to AU.



\clearpage

\clearpage








\clearpage

\end{document}